\documentclass[aps,prb,amsmath,amssymb,amstext,citeautoscript,punctuation,nofootinbib,superscriptaddress,twocolumn]{revtex4-2}
\usepackage{amsthm,mathrsfs,amsfonts,dsfont,hyperref,epsfig,braket,bm,enumerate,color,graphicx,dcolumn,charter,gensymb,comment,physics,cleveref,mathtools,natbib}
\usepackage[normalem]{ulem}
\usepackage[dvipsnames]{xcolor}
\usepackage [english]{babel}
\usepackage{graphicx}
\usepackage{dcolumn}
\usepackage{bm}
\begin{document}
\preprint{APS/123-QED}
\title{Nuclear quadrupole interaction and zero first-order Zeeman transitions of $^{167}$Er$^{3+}$ in CaWO$_4$}
\author{Lewin Marsh}
\thanks{These authors contributed equally to this work}
\affiliation{Quantum Innovation Centre (Q.InC), Agency for Science, Technology and Research (A*STAR), 2 Fusionopolis Way, Innovis \#08-03, Singapore 138634, Singapore}
\affiliation{University of Oxford, Parks Rd, Oxford OX1 3PU, United Kingdom}
\author{Yikai Yang}
\thanks{These authors contributed equally to this work}
\affiliation{Department of Engineering Science, University of Oxford, OX1 3PJ, United Kingdom}
\author{Cesare Mattiroli}
\thanks{These authors contributed equally to this work}
\affiliation{Laboratory for Quantum Magnetism, Institute of Physics, \'Ecole Polytechnique F\'ed\'erale de Lausanne (EPFL), CH-1015 Lausanne, Switzerland}
\author{Mikhael T. Sayat}
\affiliation{Quantum Innovation Centre (Q.InC), Agency for Science, Technology and Research (A*STAR), 2 Fusionopolis Way, Innovis \#08-03, Singapore 138634, Singapore}
\affiliation{Centre for Quantum Technologies, National University of Singapore, 3 Science Drive 2, Singapore 117543, Singapore}
\author{\DJ\`am Minh Tr\'i}
\affiliation{Quantum Innovation Centre (Q.InC), Agency for Science, Technology and Research (A*STAR), 2 Fusionopolis Way, Innovis \#08-03, Singapore 138634, Singapore}
\address{National University of Singapore, Science Drive 3, Singapore 117551, Singapore}
\author{Henrik M. R\o{}nnow}
\affiliation{Laboratory for Quantum Magnetism, Institute of Physics, \'Ecole Polytechnique F\'ed\'erale de Lausanne (EPFL), CH-1015 Lausanne, Switzerland}
\author{Jevon J. Longdell}
\affiliation{Department of Physics, University of Otago, Dunedin, New Zealand}
\affiliation{The Dodd–Walls Centre for Photonic and Quantum Technologies, Dunedin, New Zealand}
\author{Jian-Rui Soh}
\email{Soh\_Jian\_Rui@a-star.edu.sg}
\affiliation{Quantum Innovation Centre (Q.InC), Agency for Science, Technology and Research (A*STAR), 2 Fusionopolis Way, Innovis \#08-03, Singapore 138634, Singapore}
\affiliation{Centre for Quantum Technologies, National University of Singapore, 3 Science Drive 2, Singapore 117543, Singapore}
\affiliation{Research School of Physics, Australian National University, Canberra, ACT, 0200, Australia}
\date{\today}
\begin{abstract}
We report microwave spectroscopy of $^{167}$Er$^{3+}$ doped in CaWO$_4$ which reveals the hyperfine splitting of the erbium electronic ground state ($Z_1$, $J_\mathrm{eff.}$=15/2) induced by the $I$=7/2 nuclear spin. From spectra measured below$\sim$50 mK in magnetic fields up to 200 mT, we extract spin Hamiltonian parameters including the electron $\textbf{g}$, hyperfine $\textbf{A}$, and nuclear electric quadrupolar $\textbf{Q}$ tensors. Crucially, our analysis demonstrate unambiguously, that the previously unobserved nuclear electric quadrupolar moment is essential to reproduce the experimental data. With these refined parameters, we identify zero first-order Zeeman (ZEFOZ) transitions at zero magnetic field. Extending the analysis to finite fields, we uncover that ZEFOZ points lie either along the $c$ axis or within the $a$–$b$ plane. These results establish CaWO$_4$ as a promising host for long lifetime quantum memories.
\end{abstract}

\maketitle


\section{\label{sec:level1}Introduction}
Rare-earth-ion doped crystals are a leading platform for scalable quantum memories owing to their long coherence times, narrow optical linewidths and excellent spectral stability~\cite{thiel2011rare, GOLDNER20151,nilsson2005coherent,zhong2015optically,ranvcic2018coherence,nilsson2005coherent}. These attributes are essential for the reliable storage and retrieval of quantum information, a prerequisite for building extended quantum networks and repeaters. The quantum information is encoded within the $4f$ electronic manifold of the rare-earth ions, which is highly localized and well shielded from environmental perturbations. Within the family of rare-earth ions, erbium (Er$^{3+}$) is particularly attractive as it harbors an intra-$4f$ transition which lies within the telecommunications C-band, enabling direct interfacing with existing fiber-optic infrastructure~\cite{ourari2023indistinguishable,uysal2025spin,sayat2025polarisation}. 

\begin{figure}[b!]
\includegraphics{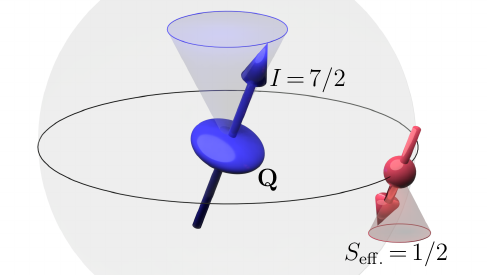}
\caption{\label{fig:Figure_1} The hyperfine interaction in $^{167}$Er$^{3+}$ between the nuclear magnetic dipole moment ($I$=7/2) and the $Z_1$ electronic ground state of the $J$=15/2 crystal electric field manifold (which can be described as an effective $S_\mathrm{eff.}$=1/2 state). The resultant 16 hyperfine energy levels can harbor clock transitions, which are also known as ZEFOZ transitions.}
\end{figure}

However, three electrons in the $4f$ shell of Er$^{3+}$ are unpaired and combine to generate a large electronic magnetic moment, rendering erbium ions highly susceptible to decoherence from external magnetic field fluctuations. This $J_\mathrm{eff}$=15/2 magnetic moment couples strongly to surrounding nuclear spins in the host crystal and to other Er$^{3+}$ magnetic dipole moments, leading to magnetic-field-dependent dephasing and reduced fidelity of stored quantum information~\cite{sayat2025polarisation}. While Er$^{3+}$–Er$^{3+}$ dipolar interactions can be mitigated by using low dopant concentrations (below 100 ppm)~\cite{matsuura2025explorationoptimalhyperfinetransitions}, decoherence arising from nuclear spin bath remains a significant limitation. Therefore, the identification of host crystals with a low natural abundance of nuclear spins is crucial for realizing the full potential of erbium-based quantum memories.

\begin{figure}[b!]
    \centering
    \includegraphics[width=0.99\linewidth]{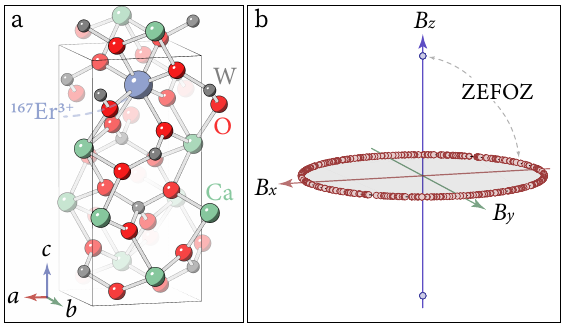}
    \caption{\textbf{(a)} The crystal structure of CaWO$_4$ with Er$^{3+}$ dopant ion residing on the Ca$^{2+}$ site. \textbf{(b)} Magnetic field distribution of ZEFOZ points based on our three-dimensional \textit{\textbf{B}} field search. The points lie either along the crystal $c$-axis or within the $a$-$b$ plane.}
    \label{fig:Figure_2}
\end{figure}

Recently, calcium tungstate (CaWO$_4$) has emerged as an especially promising solid-state host for erbium ions [Fig.~\ref{fig:Figure_2}\textbf{a}], exhibiting exceptionally long spin coherence times and narrow optical linewidths~\cite{le2021twenty,yang2023spectroscopic,becker2025spectroscopic,an2025optical,tiranov2025sub,li_realization_2025}. This performance stems from the relatively quiet magnetic environment of CaWO$_4$, with nuclear-active isotopes limited to $^{183}$W (14.3\%, $I$=1/2), $^{17}$O (0.03\%, $I$=-5/2), and $^{43}$Ca (0.135\%, $I$=-7/2). Nonetheless, even this dilute nuclear spin bath generates low-frequency magnetic field fluctuations that couple to the large Er$^{3+}$ moment, ultimately imposing an upper limit on coherence times and reducing the fidelity of stored quantum information. 

An effective strategy to mitigate decoherence from such magnetic fluctuations is to operate at zero first-order Zeeman (ZEFOZ) conditions, where the transition frequency is invariant to linear magnetic fluctuations~\cite{judd1961theory}. These ZEFOZ transitions provide quantum states protection against magnetic noise, as the gradient of the transition frequency with respect to magnetic field is zero. This technique has proven highly effective in other rare-earth systems, such as Eu$^{3+}$:Y$_2$SiO$_5$ \cite{longdell2006characterization}, where ZEFOZ transitions have extended spin coherence times to over six hours \cite{zhong2015optically,wang2025nuclear}. These transitions reside within the hyperfine manifold of the electronic ground state, which is present only in isotopes with a non-zero nuclear magnetic dipole moment.

Only one stable isotope of erbium has a nuclear spin
, namely $^{167}$Er with $I$$=$$7/2$. Through hyperfine coupling, this nuclear moment, $I$, lifts the degeneracy of the lowest crystal-field level ($Z_1$) of the $J_\mathrm{eff} = 15/2$ ground-state multiplet, resulting in 16 distinct hyperfine sublevels~\cite{rakonjac2020long}. These hyperfine levels support 120 magnetic-dipole-allowed transitions, offering a rich landscape for ZEFOZ transitions.

Despite this potential, ZEFOZ conditions in $^{167}$Er$^{3+}$:CaWO$_4$ have not been systematically explored. Candidate ZEFOZ points could, in principle, be inferred from previous reports of the $\textbf{g}$ and $\textbf{A}$ tensors of the hyperfine spin Hamiltonian of $^{167}$Er$^{3+}$ in CaWO$_4$~\cite{antipin1968paramagnetic}. However, those electron spin resonance measurements were conducted predominantly at high magnetic fields and did not extend into the zero-field regime. As such, the nuclear electric quadrupole interaction, which is expected to be substantial for Er$^{3+}$ ions substituting Ca$^{2+}$ sites, was neglected. The residual electric field gradients caused by the necessary charge compensation at these sites produce a strong quadrupolar interaction that must be included to accurately model the spin Hamiltonian and predict ZEFOZ transitions, particularly those near zero field.

In this work, we present detailed microwave spectroscopy measurements of $^{167}$Er$^{3+}$ dopants in CaWO$_4$ at milliKelvin temperatures and applied magnetic fields up to 200\,mT. The measured spectra reveals the hyperfine structure of the electronic ground state of erbium, including a zero-field splitting. Crucially, we find that a previously neglected nuclear electric quadrupole interaction is essential to accurately reproduce the measured spectra. Incorporating this term into the spin Hamiltonian, we identify candidate ZEFOZ transitions that occur at zero-field.

Extending our analysis to non-zero fields we found ZEFOZ points both along the $c$ axis and as rings in the $a$-$b$ plane [Fig.~\ref{fig:Figure_2}\textbf{b}]. 
%
Our calculations indicate that at field strengths of $\sim$2\,T perpendicular to the $c$ axis, significantly enhanced coherence times are possible. These findings establish CaWO$_4$ as a promising platform for erbium-based quantum memories, combining long spin coherence with telecom-band optical access.

\section{Methods}
Single crystalline CaWO$_4$ was grown by the hybrid Czochralski method and doped with naturally-occurring erbium at a target concentration of 100 ppm. After growth, the crystal was oriented by x-ray Laue diffraction and cut in 5$\times$5$\times$1mm$^{3}$ plates along the tetragonal $a$ and $c$ axes of the $I4_1/a$ crystal unit cell. The samples were mechanically polished to an epitaxial finish so as to reduce microwave scattering losses.

\subsection{Microwave spectroscopy}
For microwave spectroscopy, the single-crystal samples were glued with GE varnish onto a copper coplanar waveguide (CPW) such that the crystal $c$ axis was aligned along the length of the CPW central conductor. The CPW was mounted on the cold stage of a dilution refrigerator with a base temperature $T\sim$50 mK. The scattering parameter $S_{21}$ was measured by a vector network analyzer (VNA). Thermal noise on the input line is suppressed by a series of attenuators (4K stage: 20 dB, still: 10 dB, cold plate: 10 dB, mixing chamber: 20 dB). The output signal is amplified by 40 dB cryogenic low-noise amplifier, and by a second 30 dB amplification stage at room temperature. Back-propagating noise from the amplifiers is screened by a circulator. A superconducting vector magnet provided magnetic fields up to 1\,T in arbitrary orientations relative to the tetragonal crystal axes. For each field scan, the trace measured in the largest applied field was chosen as a reference to normalize the others $S_{21}^{\text{norm}}(B)=S_{21}(B)/S_{21}(B_{\text{ref}})$ to suppress the background.

\subsection{Theoretical model}
The electronic ground state of $^{167}$Er$^{3+}$ in CaWO$_4$ arises from the lowest crystal electric field doublet ($Z_1$) of the $J$=15/2 manifold, which can be described by an effective spin-1/2 coupled to the $^{167}$Er nuclear spin.
The spin Hamiltonian can be expressed by,
\begin{equation}
    \hat{H}=\mu_\mathrm{B} \textbf{B}\cdot\textbf{g}\cdot \hat{\textbf{S}}-\mu_\mathrm{n}g_\mathrm{n}\textbf{B}\cdot\hat{\textbf{I}}+\hat{\textbf{I}}\cdot\textbf{A}\cdot\hat{\textbf{S}}+\hat{\textbf{I}}\cdot\textbf{Q}\cdot\hat{\textbf{I}}
    \label{eq:spin_hamiltonian}
\end{equation}
where $\hat{\textbf{S}}$ and $\hat{\textbf{I}}$ are the electronic and nuclear spin operators, respectively. Here $\mu_\mathrm{B}$ is the Bohr magneton, 
$\textbf{g}$ is the anisotropic electronic
$g$ tensor, $\textbf{A}$ is the hyperfine coupling tensor, $\textbf{Q}$ is the nuclear electric quadrupole tensor and $g_\mathrm{n}$ is the nuclear
$g$ factor of $^{167}$Er.

The first term describes the electronic Zeeman interaction, the second the nuclear Zeeman interaction, the third the hyperfine coupling between the nuclear and electronic moments, and the fourth the nuclear electric quadrupole interaction arising from local electric field gradients at the Er$^{3+}$ site. The latter is expected to be substantial because Er$^{3+}$ substitutes for Ca$^{2+}$ with strong crystal field gradients. Owing to the $S_4$ symmetry of the substitutional site, $\textbf{g}$, $\textbf{A}$ and $\textbf{Q}$ are constrained to be diagonal with the first two diagonal elements equal. Furthermore, the $\textbf{Q}$ tensor possess $z^2$ symmetry and is hence traceless, where $Q_{xx}$=$Q_{yy}$=-$Q_{zz}/2$.

The $\textbf{g}$, $\textbf{A}$, and $\textbf{Q}$ tensors were extracted by globally fitting the microwave spectra across multiple magnetic field orientations using a nonlinear least-squares routine. Simulations of the spectra was performed by numerically diagonalizing $\hat{H}$ and computing the magnetic susceptibility $\chi_{xx}$ as a function of field, based on the experimental configuration.
 
\subsection{Identification of ZEFOZ points}
To identify ZEFOZ transitions, we constructed the spin Hamiltonian [Eq.~\eqref{eq:spin_hamiltonian}] using the refined $\textbf{g}$, $\textbf{A}$, and $\textbf{Q}$ tensor parameters. The Hamiltonian was numerically diagonalized to obtain the eigenenergies $E_n(\textbf{B})$ and corresponding eigenstates, at any given applied magnetic field $\textbf{B}$. As described earlier, diagonalization yields sixteen hyperfine eigenstates of the $Z_1$ ground-state manifold, labeled $|0\rangle, |1\rangle, \dots, |15\rangle$, ordered from the lowest to highest energy at zero magnetic field. These states give rise to 120 allowed magnetic-dipole transitions, denoted by $f_{i\,j}$ with $i, j = 0,1,\dots,15$ and $i<j$.

The transition frequencies were defined as $f_{i\,j}(\textbf{B})$ = $[E_j(\textbf{B})-E_i(\textbf{B})]/{h}$,
with their magnetic sensitivities quantified by the gradient, $\textbf{S}_1^{(i\,j)}$=$\nabla_\textbf{B} f_{i\,j}(\textbf{B})$,
and the residual quadratic coupling described by the curvature, $\textbf{S}_2^{(i\,j)}$ = $\tfrac{1}{2}\nabla^2_\textbf{B} f_{i\,j}(\textbf{B})$. In terms of these quantities, the hyperfine coherence time is estimated as,
\begin{equation}\label{eq:T2hyp}
\frac{1}{\pi T_2^\mathrm{hyp}} = \textbf{S}^{(i\,j)}_1\cdot\Delta\textbf{B}
+ \Delta\textbf{B}\cdot\textbf{S}^{(i\,j)}_2\cdot\Delta\textbf{B},
\end{equation}
where $\Delta\textbf{B}$ characterizes the ambient magnetic noise (e.g. from the nuclear spin bath of the host crystal)\footnote{For the sake of brevity, we will drop the indices, $i$, $j$.}~\cite{longdell2006characterization,wang2023hyperfine,McAuslan2012_Zefoz}. 

ZEFOZ points are formally defined by the condition $|\textbf{S}_1|$=0, at which the transition frequency becomes first-order insensitive to magnetic field fluctuations~\cite{longdell2006characterization,zhong2015optically,wang2023hyperfine,McAuslan2012_Zefoz}. In this regime, only second-order (quadratic) couplings remain, leading to a strong suppression of dephasing and extended coherence times $T_2^\mathrm{hyp}$ provided that both the magnitude of the ambient noise $\Delta\textbf{B}$ and the curvature $\textbf{S}_2$ are also sufficiently small~\cite{longdell2006characterization,zhong2015optically,wang2023hyperfine,McAuslan2012_Zefoz}. 


\begin{figure}[t!]
    \centering
    \includegraphics[width=\linewidth]{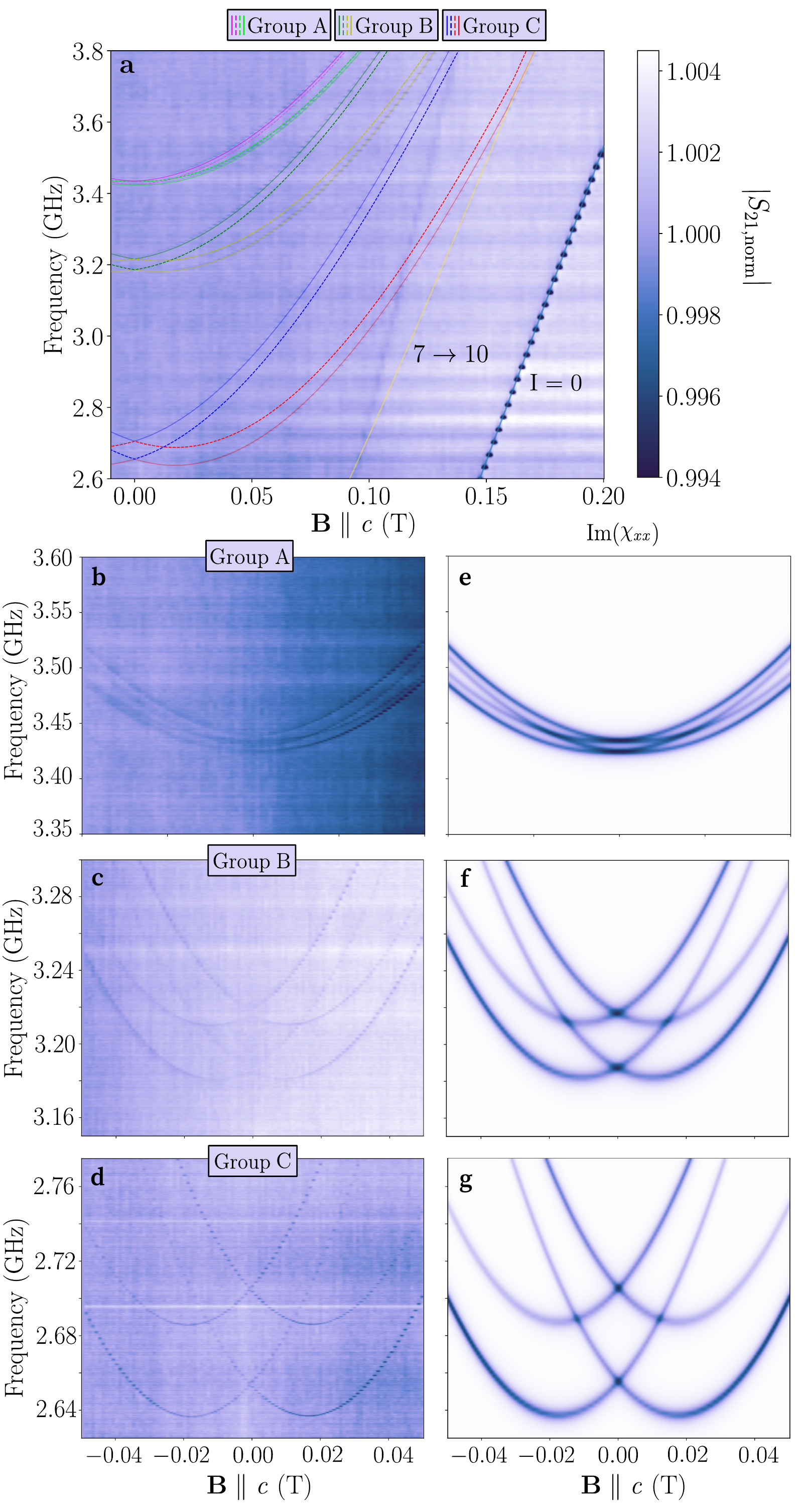}
    \caption{\label{fig:Figure_3} \textbf{(a)} Normalized $S_{21}$ transmission as a function of field along the crystal $c$ axis, overlaid with the spectrum calculated from the refined spin Hamiltonian parameters. To highlight the zero-field splitting due to the nuclear electric quadrupolar moment of $^{167}$Er$^{3+}$, we plot in panels \textbf{(b)}--\textbf{(d)} the measured spectrum in a smaller frequency-field window. Panels \textbf{(e)}--\textbf{(g)} plots the calculated susceptibilities, which agree well with the measurements. We observed all visible transitions predicted by susceptibility calculations, apart from the line cutting across group C, which we are not able to explain.}
\end{figure}

 To identify such ZEFOZ candidates and estimate the corresponding $T_2^\mathrm{hyp}$, we computed the gradient ($\textbf{S}_1$) and curvature ($\textbf{S}_2$) using perturbation theory~\cite{longdell2006characterization,McAuslan2012_Zefoz} 
 while the ambient noise $\Delta\textbf{B}$ was obtained via a Monte-Carlo simulation of the magnetic noise experienced by an Er ion embedded within a 8$\times$8$\times$8  supercell of CaWO$_4$ (See Appendix for details). The search over all 120 possible transitions was performed not only at zero-magnetic field ($\textbf{B}$=0), but also at arbitrary magnetic field vectors, $\textbf{B}$=$(B_x,B_y,B_z)$, for transitions where the gradient satisfies the $|\textbf{S}_1|$$<$$10^{-8}$\,GHz\,T$^{-1}$ threshold\footnote{While ZEFOZ points are formally defined by $|\textbf{S}_1|$=0, in practice we identify candidates when $|\textbf{S}_1|$$<$$10^{-8}$\,GHz\,T$^{-1}$, which is effectively zero within numerical precision.}. 
 
 For finite fields, the search was initialized on a three-dimensional grid spanning 0-20\,T, defined with respect to the crystallographic $a$, $b$, and $c$ axes [See Fig.~\ref{fig:Figure_2}\textbf{a}], with a step size of 2.5\,mT. From each grid point, we performed a three-dimensional search using the Newton-Raphson optimization method~\cite{longdell2006characterization,McAuslan2012_Zefoz}, up to a maximum search bound of 200\,T. 

\section{Results}
Figure~\ref{fig:Figure_3}\textbf{a} plots the normalized $S_{21}$ transmission measurements with the magnetic field applied along the crystal $c$ axis. Several absorption lines are observed, arising from transitions between eigenstates of Eq.~\ref{eq:spin_hamiltonian}. The dominant feature is a strong mode associated with the nuclear-inactive ($I$=0) erbium isotopes, which produce a single transition [See Fig.~\ref{fig:Figure_3}\textbf{a}]. 

In addition, weaker features originate from the hyperfine manifold of nuclear-active isotope $^{167}$Er$^{3+}$ ($I$=$\frac{7}{2}$, 23\% natural abundance). Although this manifold supports 120 possible transitions, most lie outside the accessible frequency-field window or have vanishing matrix elements in our experimental configuration. Using the spin Hamiltonian parameters from Ref.~\cite{antipin1968paramagnetic} as a starting point, we calculated the magnetic susceptibility $\chi_{xx}$ of $^{167}$Er$^{3+}$ to ascertain the subset of transitions expected within the frequency-field window of Fig.~\ref{fig:Figure_3}\textbf{a}. These transitions cluster into three groups: Group A  $|6\rangle$$\rightarrow$$|11\rangle$, $|5\rangle$$\rightarrow$$|12\rangle$, $|4\rangle$$\rightarrow$$|9\rangle$, $|3\rangle$$\rightarrow$$|10\rangle$; Group B $|4\rangle$$\rightarrow$$|13\rangle$, $|3\rangle$$\rightarrow$$|14\rangle$, $|2\rangle$$\rightarrow$$|11\rangle$, $|1\rangle$$\rightarrow$$|12\rangle$; Group C $|2\rangle$$\rightarrow$$|15\rangle$, $|1\rangle$$\rightarrow$$|15\rangle$, $|0\rangle$$\rightarrow$$|13\rangle$, $|0\rangle$$\rightarrow$$|14\rangle$. 
 
The calculated spectrum based on the parameters in Ref.~\cite{antipin1968paramagnetic}, which neglects the nuclear electric quadrupolar interaction $\textbf{Q}$, fails to reproduce the observed zero-field splittings. As highlighted in Figs.~\ref{fig:Figure_3}\textbf{b}, \textbf{c}, groups B and C exhibit zero-field splittings of $\sim$30 MHz and $\sim$50\,MHz, respectively, which should be absent if $\textbf{Q}$ is zero.

Therefore, we refined the spin Hamiltonian parameters (Eq.~\ref{eq:spin_hamiltonian}) against our measured data. Crucially, we found that the spin Hamiltonian can only reproduce the data if we include a non-zero electric nuclear quadrupole moment, $\textbf{Q}$. Indeed, the calculated spectra show excellent agreement with experiment [Fig.~\ref{fig:Figure_3}\textbf{a}], and the corresponding susceptibilities [Figs.~\ref{fig:Figure_3}\textbf{e}--\textbf{f}] match the measured transmission [Figs.~\ref{fig:Figure_3}\textbf{b}--\textbf{d}]. The parameters obtained from our fit are as follows, 
\begin{equation*}
\begin{split}
    \boldsymbol{g}&=\left[\begin{array}{ccc}
     8.3(1) & 0 & 0 \\
        0 & 8.3(1) & 0\\
      0 & 0 & 1.262(3)  
    \end{array}\right],\\
        \boldsymbol{A}&=\left[\begin{array}{ccc}
     -871.09(4) & 0 & 0 \\
        0 & -871.09(4) & 0\\
      0 & 0 & -128.3(2)  
    \end{array}\right]\mathrm{MHz},\\
    \boldsymbol{Q}&=\left[\begin{array}{ccc}
     1.68(2) & 0 & 0 \\
        0 & 1.68(2) & 0\\
      0 & 0 & -3.36(4)  
    \end{array}\right]\mathrm{MHz}.\\
\end{split}
\end{equation*}

Notably, our $\textbf{g}$ and $\textbf{A}$ parameters are broadly consistent with the previously reported values in Ref.~\cite{antipin1968paramagnetic}. However, given that the measurements by Antipin \textit{et al.}~\cite{antipin1968paramagnetic} were performed at relatively high magnetic field strengths -- where the effect of the nuclear quadrupole interaction is negligible -- their study was insensitive to the presence of $\textbf{Q}$. Indeed, many subsequent investigations of the hyperfine levels of erbium in CaWO$_4$ have likewise omitted $\textbf{Q}$~\cite{MASON1968260,li_realization_2025, le2021twenty,CHANELIERE2024120647,ourari2023indistinguishable, wang2024monthlonglifetimemicrowavespectralholes,Rancic_2022,bertaina_rare-earth_2007}. In contrast, by probing down to zero field, we are able to resolve the lifting of degeneracies in specific transitions, providing direct evidence of the NQI.

\subsection{ZEFOZ points at zero-field}

Incorporating the nuclear quadrupole interaction $\mathbf{Q}$ into the spin Hamiltonian is essential for identifying ZEFOZ transitions, particularly those occurring at zero magnetic field. We utilized second-order perturbation theory to obtain $\textbf{S}_1$ and $\textbf{S}_2$ at zero-field, due to the fact that many of the states fell into degenerate pairs. Among the 120 hyperfine transitions at $\textbf{B}$=0\,T, only one exhibits a sufficiently small gradient to satisfy the ZEFOZ condition with the gradient $|\textbf{S}_1|$$<$$10^{-8}$\,GHz\,T$^{-1}$. 

This zero-field ZEFOZ corresponds to the $|0\rangle$$\rightarrow$$|15\rangle$ transition at $f$=3.4844\,GHz [See Table~\ref{tab:0field}]. The associated curvature, however, is comparatively large ($|\mathbf{S}_2|=1.278$$\times$$10^{5}$ GHz\,T$^{-2}$), which is expected to limit coherence times. For this reason, we also considered additional transitions that, while not strictly ZEFOZ, possess sufficiently small gradients and more favorable curvature.

We append to the list in Table~\ref{tab:0field}, six near-ZEFOZ transitions at $\textbf{B}$=0\,T, ordered by increasing $|\textbf{S}_1|$, along with their initial and final eigenstates, transition frequencies, and $|\textbf{S}_2|$ values. These occur as three degenerate pairs at $f$=2.3328, 3.0272, and 3.3759 GHz, each with $|\mathbf{S}_1|$$\sim$$10^{-3}$\,GHz\,T$^{-1}$, which is much larger than in the $|0\rangle$$\rightarrow$$|15\rangle$ case. Nevertheless, their curvatures are significantly smaller [Table~\ref{tab:0field}], suggesting that these transitions may provide greater robustness against magnetic-field fluctuations despite their higher first-order sensitivity.

In order to estimate the coherence times of these seven candidate transitions, we evaluated the magnetic-field fluctuations at $\textbf{B}$=0\,T using a Monte Carlo simulation, as described earlier in the Methods section. We find that both the CaWO$_4$ host and the Er$^{3+}$ spins contribute fluctuations of approximately $7$\,$\mu$T at zero field. 
At this level of magnetic fluctuations, the best-performing transition is still the only true ZEFOZ transition, which is $|0\rangle$$\rightarrow$$|15\rangle$. It has an estimated $T_2^\mathrm{hyp}$$\approx$$12.7\,\mu$s, far outperforming the other transitions. Although our data shows that $|0\rangle$$\rightarrow$$|15\rangle$ has very low $\chi_{xx}$, it is strongly visible in $\chi_{zz}$, which makes it a good zero-field ZEFOZ candidate.

\begin{table}[t!]
\caption{ZEFOZ and near-ZEFOZ transitions within the ground state ($J_\mathrm{eff}$=15/2, $Z_1$) hyperfine manifold of $^{167}$Er$^{3+}$ in CaWO$_4$ at \textbf{B}=0\,T.}
\label{tab:0field}
\begin{ruledtabular}
\begin{tabular}{cccc}
 $|i\rangle$$\rightarrow$$|j\rangle$  & $f$ & $|\textbf{S}_1|$ ($10^{-3}$& $|\textbf{S}_2|$ ($10^3$ \\
  & (GHz)& GHz\,T$^{-1}$)& GHz\,T$^{-2}$)\\[1ex] 
 \hline \hline   $|0\rangle$$\rightarrow$$|15\rangle$ & 3.4844 &0& 127.8 \\  [1ex]
 \hline
 $|1\rangle$$\rightarrow$$|13\rangle$ & 3.3759 & 2.467 & 41.38 \\  
 $|2\rangle$$\rightarrow$$|14\rangle$ & 3.3759 & 2.467 & 49.43 \\  [1ex]
 \hline
 $|3\rangle$$\rightarrow$$|11\rangle$ & 3.0272 & 4.933 & 7.482 \\  
 $|4\rangle$$\rightarrow$$|12\rangle$ & 3.0272 & 4.933 & 7.482 \\  [1ex]
 \hline
 $|5\rangle$$\rightarrow$$|9\rangle$ & 2.3328 & 7.400 & 2.526 \\  
 $|6\rangle$$\rightarrow$$|10\rangle$ & 2.3328 & 7.400 & 2.526 \\  [1ex]
   \end{tabular}
\end{ruledtabular}
\end{table}

\subsection{ZEFOZ points in finite magnetic field}
Based on our Newton-Raphson search in finite $\textbf{B}$ fields for hyperfine transitions satisfying the condition $|\textbf{S}_1|$$<$10$^{-8}$GHz\,T$^{-1}$, we found a total of 163 unique ZEFOZ points. Each point was recorded at least eleven times, confirming that the solutions are robust and unique. To understand their magnetic-field orientation distribution, we project all the ZEFOZ points onto a sphere in Fig.~\ref{fig:Figure_2}\textbf{b}. Interestingly, we found that the ZEFOZ points all cluster either with magnetic field aligned along the crystal $c$ axis (blue) or within the $a$-$b$ plane (red). 



ZEFOZ points along the $c$ axis arise when the $z$-component of the transition-frequency gradient with respect to the magnetic field is zero, while cylindrical symmetry ensures the transverse components are zero. In contrast, in-plane ZEFOZ points occur when the radial component of the gradient is zero due to mirror symmetry of the spin Hamiltonian in the $a$–$b$ plane. Cylindrical symmetry further implies that any such solution can be rotated by an arbitrary angle about the $c$ axis to generate another ZEFOZ point, yielding continuous concentric rings of in-plane solutions. This ring-like manifold directly reflects the axial ($U(1)$) symmetry of the Hamiltonian in Eq.~\ref{eq:spin_hamiltonian}, since rotations of an in-plane magnetic field about the $c$ axis leave the Hamiltonian invariant and produce families of symmetry-equivalent ZEFOZ orientations.

\begin{table}[b!]
\caption{\label{tab:ZEFOZS2} Several of the best ZEFOZ transitions within the ground state ($J_\mathrm{eff}$=15/2, $Z_1$) hyperfine manifold of $^{167}$Er$^{3+}$ in CaWO$_4$ at finite magnetic fields, \textbf{B}. The ZEFOZ transitions perpendicular to the $c$-axis outperform those along the $c$-axis.}
\begin{ruledtabular}
\begin{tabular}{c c c c} 
  $|i\rangle$$\rightarrow$$|j\rangle$  & $\textbf{B}$\,(T) & $f$\,(MHz)& $|\textbf{S}_2|$\,(GHz\,T$^{-2}$)\\ [1ex]
 \hline
 $|5\rangle$$\rightarrow$$|6\rangle$ &(1.569,0,0) & 449.6  & 2.04 \\  
 $|13\rangle$$\rightarrow$$|14\rangle$ &(1.587,0,0) & 441.8  & 1.93 \\  [1ex]
 \hline
 $|14\rangle$$\rightarrow$$|15\rangle$ &(1.980,0,0) & 445.6  & 1.24 \\  
 $|6\rangle$$\rightarrow$$|7\rangle$ & (1.997,0,0) & 455.4  & 1.26 \\ [1ex]
 \hline
 $|12\rangle$$\rightarrow$$|14\rangle$ & (0,0,4.208) & 77.40  & 1240 \\ 
\end{tabular}
\end{ruledtabular}
\end{table}

\begin{figure}[t!]
    \centering
    \includegraphics[width=0.9\linewidth]{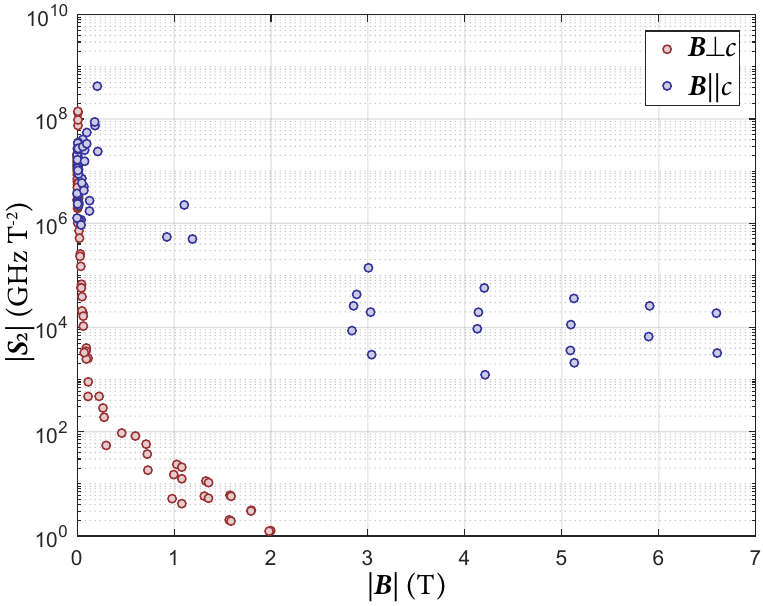}
    \caption{The distribution of the ZEFOZ points as function of $|\mathbf{S}_2|$ and magnetic field strength, parallel and perpendicular to the crystal $c$ axis.}
    \label{fig:Figure_4}
\end{figure}

The distribution of these ZEFOZ points as a function of magnetic field strength is plotted in Fig.~\ref{fig:Figure_4}, together with the associated curvature $|\mathbf{S}_2|$, which sets the residual quadratic coupling to magnetic noise. Notably, all of the ZEFOZ points within the $a$-$b$ plane occur below a field of 2\,T and those which lie along the crystal $c$ axis are below 7\,T in magnetic field, which are well within the reach of a standard cryomagnet. 

Several of the most favorable candidates, characterized by the smallest $|\mathbf{S}_2|$, are summarized in Table~\ref{tab:ZEFOZS2}, along with the associated magnetic field location, and transition frequencies. In particular, the $|14\rangle$$\rightarrow$$|15\rangle$ and $|6\rangle$$\rightarrow$$|7\rangle$ transitions near $\mathbf{B}$$\sim$2\,T along the crystal $a$ axis exhibits the smallest $|\mathbf{S}_2|$ values. By contrast, the $|12\rangle$$\rightarrow$$|14\rangle$ transition at $\mathbf{B}$=4.2\,T along the $c$ axis displays a $|\mathbf{S}_2|$ more than three orders of magnitude larger than those with field perpendicular to the $c$ axis.


To quantitatively assess the $T_2^\mathrm{hyp}$ of these ZEFOZ points, we performed Monte Carlo simulations of the ambient magnetic noise ($\Delta B$) for fields $\textbf{B}$$\parallel$$a$ $\sim$2\,T and $\textbf{B}$$\parallel$$c$ $\sim$4\,T. The host lattice contributes $\sim$7\,$\mu$T, with an additional $\sim$2\,$\mu$T  arising from Er$^{3+}$ electron spins. 
We find that ZEFOZ transitions perpendicular to the $c$ axis yield coherence times exceeding 3\,s, three orders of magnitude longer than those along the $c$ axis. 

\begin{figure}
    \centering
    \includegraphics[width=\linewidth]{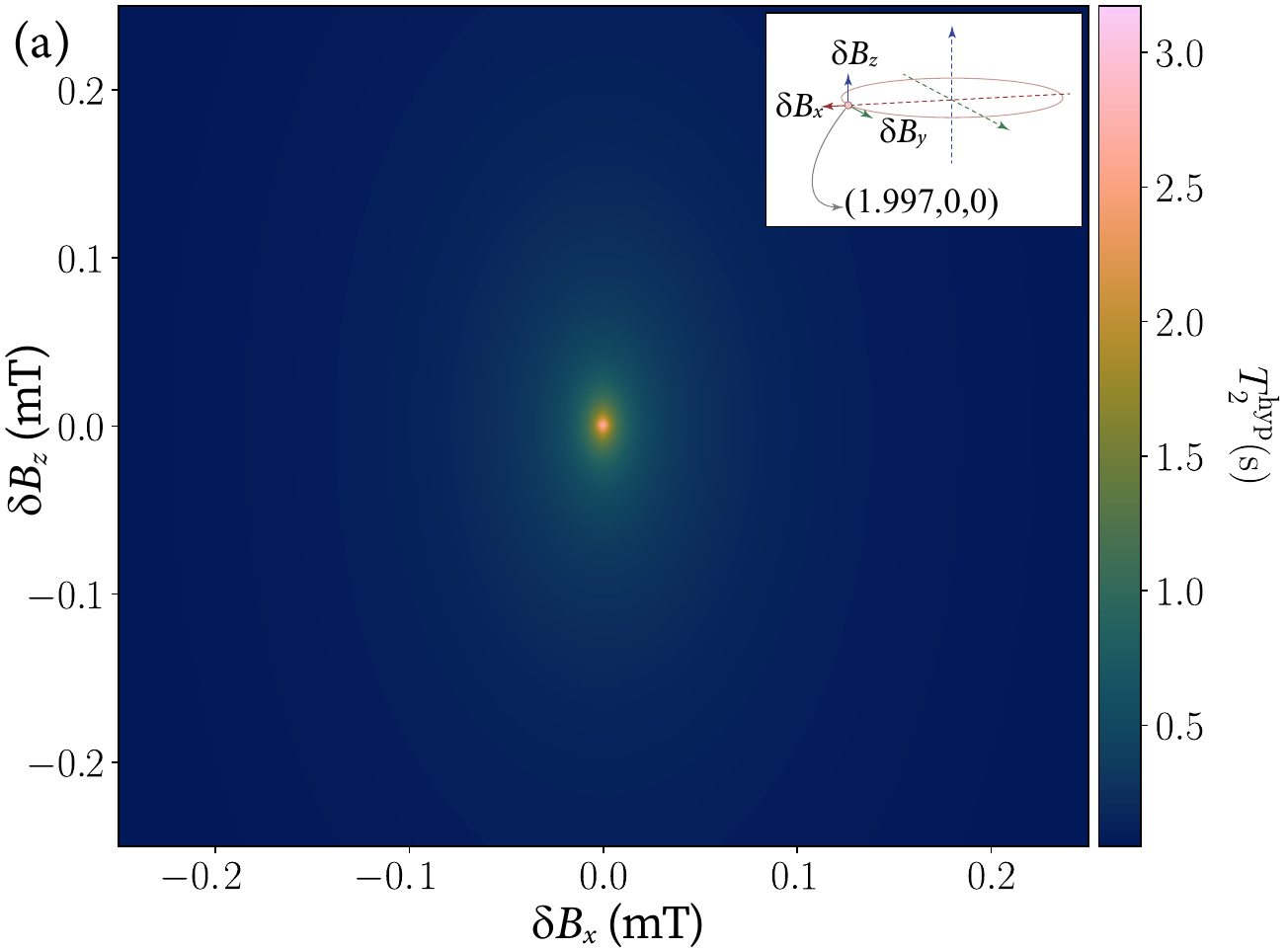}
    \includegraphics[width=\linewidth]{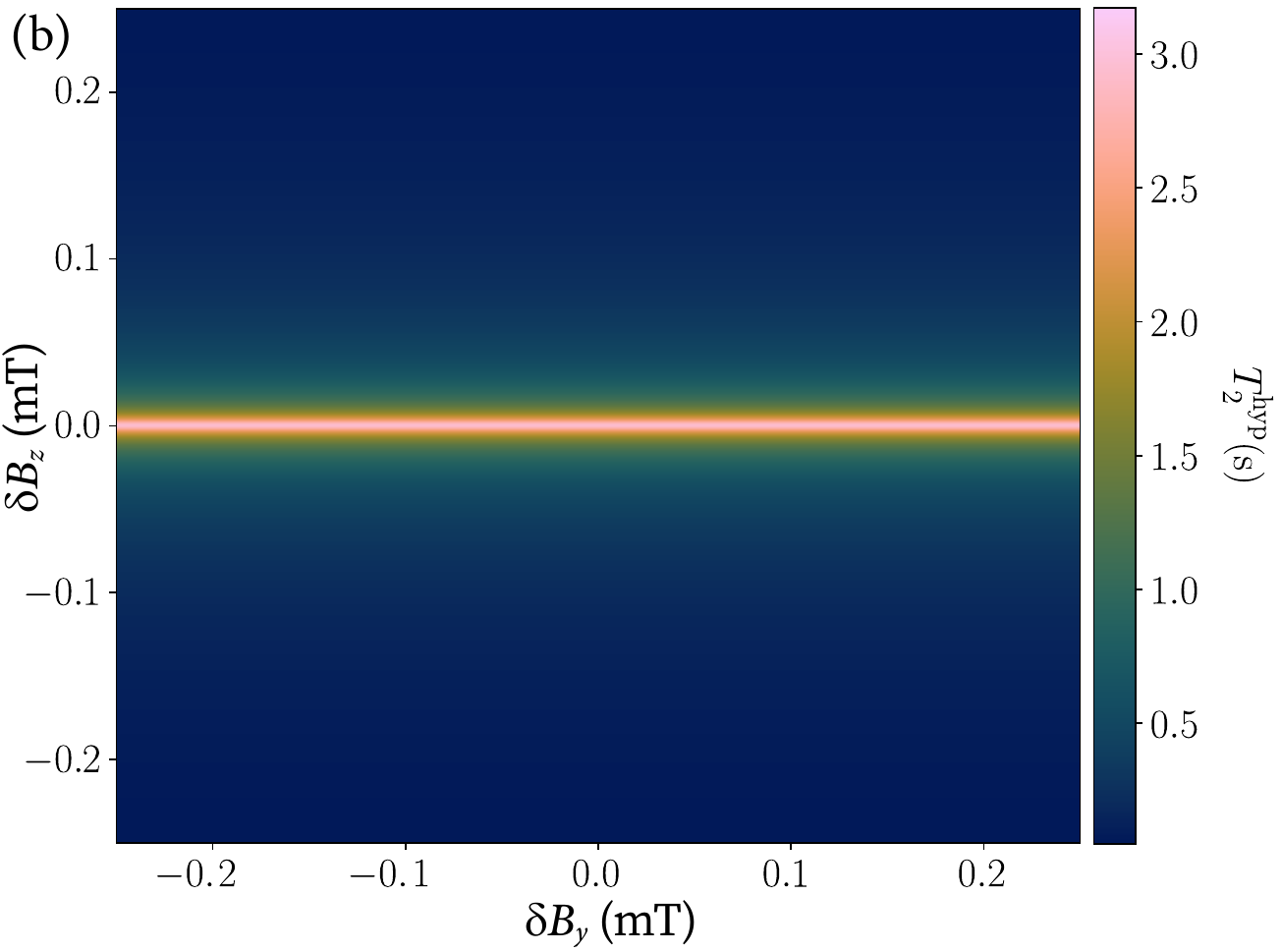}
    \caption{The stability of the  $|14\rangle$$\rightarrow$$|15\rangle$ transition at $\textbf{B}$=(1.980,0,0), with respect to magnetic fluctuations in the (a) $\delta B_x$-$\delta B_z$ and (b) $\delta B_y$-$\delta B_z$ directions, respectively. (insert) the definition of magnetic field deviations $\delta B_x$, $\delta B_y$ and $\delta B_z$ with respect to the ZEFOZ point. (a) The heatmap indicates that the coherence time is more resilient against misalignment in the \textit{z} direction. (b) The coherence time is also invariant with respect to the misalignment along the ring.}
    \label{fig:figure5}
\end{figure}

To assess the robustness of ZEFOZ operation against magnetic-field misalignment, we computed $T_2^\mathrm{hyp}$ for the $|14\rangle$$\rightarrow$$|15\rangle$ ZEFOZ candidate as a function of magnetic field fluctuations away from $\textbf{B}$=(1.980,0,0). Field deviations $\delta B_x$, $\delta B_y$ and $\delta B_z$ were introduced on a grid within $\pm$0.2\,mT around the ZEFOZ point. The results, shown in Fig.~\ref{fig:figure5}\textbf{a}, reveal that the coherence is fairly resilient to fluctuations along the $c$ axis, whereas even small deviations in field magnitude along the $x$ axis reduce $T_2^\mathrm{hyp}$ by an order of magnitude. By contrast, fluctuations along the ring ($\delta B_y$) leave $T_2^\mathrm{hyp}$ unchanged [Fig.~\ref{fig:figure5}\textbf{b}]. Thus, precise control of magnetic field strength is more critical than angular alignment, further supporting the experimental viability of ZEFOZ points perpendicular to the $c$ axis.


\section{Conclusion}
In this manuscript, we have demonstrated the suitability of $^{167}$Er$^{3+}$:CaWO$_4$ as a platform for spin-wave storage with long coherence times. Our microwave spectroscopy measurements on a sample of calcium tungstate doped with erbium ions at a concentration of 100 ppm at $T\sim$50\,mK, reveals well-resolved hyperfine absorption lines. These are accurately reproduced by our spin Hamiltonian only when a finite nuclear quadrupole interaction was included. Using these parameters, we identified ZEFOZ points both at zero-applied field, and at finite fields, either along the crystal $c$ axis or in rings perpendicular to it. Several magnetic field configuration support coherence times on the order of seconds, with the longest at 1.98\,T perpendicular to the crystal $c$-axis. These exceptional coherence times are ultimately due to the simple and quiet crystal structure of the CaWO$_4$ host lattice, underscoring its promise for rare-earth-based quantum memories.

\section*{Data Availability}

The data that support the findings of this study are available from the corresponding author upon reasonable request.

\begin{acknowledgments}
The authors wish to thank N. Kukharchyk, G. Mair, R. Ahlefeldt, M. Sellars, N. Chilton for helpful discussions. This research is supported by A*STAR under Project No. C230917009, Q.InC Strategic Research and Translational Thrust and the MTC Young Investigator Research Grant (Award \# M24N8c0110). 
\end{acknowledgments}

\appendix
\section{Computational details}
\label{Sec:comput}

 For finite fields, the exact first ($\textbf{S}_1$) and second-order ($\textbf{S}_2$) derivatives of the eigenenergies with respect to the magnetic field were calculated using the Hellmann-Feynman approach. In zero field, due to the degeneracies in the eigenstates, we were unable to use Hellman-Feynman theorem. Instead, we first numerically diagonalized the full spin Hamiltonian to obtain the eigenenergies $E_n(\mathbf{B})$ and eigenstates for any applied magnetic field $\mathbf{B}$~\cite{longdell2006characterization,McAuslan2012_Zefoz}. The magnetic-field gradient ($\mathbf{S}_1$) and curvature ($\mathbf{S}_2$) of each eigenenergy were then evaluated using second-order perturbation theory, applied directly to the numerically exact eigenstates~\cite{longdell2006characterization,McAuslan2012_Zefoz}.

To estimate the ambient magnetic noise $\Delta \textbf{B}$ experienced by the erbium ions, we performed the simulations on an ensemble of 1000 independent copies of a CaWO$_4$ superlattice comprised of 8$\times$8$\times$8 unit cells. For each configuration, we randomized the orientation of all nuclear spins in the host and used the Boltzmann distribution on the electron spin at finite fields. The nuclear active species we considered are: 
%
\begin{itemize}
    \item Er$^{3+}$(doped at 100\,ppm) electron spins, where $\boldsymbol{\mu}_e=\mu_b\boldsymbol{g}_e\cdot\boldsymbol{S}$
    \item $^{167}$Er$^{3+}$(23.0\% abundance), $|\boldsymbol{\mu}|=0.5638\mu_n$
    \item $^{183}$W$^{6+}$(14.3\% abundance), $|\boldsymbol{\mu}|=0.11778476\mu_n$
    \item $^{43}$Ca$^{2+}$(0.135\% abundance), $|\boldsymbol{\mu}|=1.3176\mu_n$
    \item $^{17}$O$^{2-}$(0.038\% abundance), $|\boldsymbol{\mu}|=1.893\mu_n$
\end{itemize}
\bibliography{apssamp}
\end{document}